\documentclass[fleqn,usenatbib]{mnras}

\usepackage{newtxtext,newtxmath}

\usepackage[T1]{fontenc}
\usepackage{ae,aecompl}

\usepackage{graphicx}	
\usepackage{amsmath}	
\usepackage{amssymb}	

\graphicspath{{figures/}}

\title[Impact of Peculiar Velocity on $H_0$]{The Impact of Peculiar Velocities on the Estimation of the Hubble Constant from Gravitational Wave Standard Sirens}

\author[Constantina Nicolaou et al.]{
Constantina Nicolaou,$^{1}$\thanks{E-mail: constantina.nicolaou.17@ucl.ac.uk}
Ofer Lahav,$^{1}$
Pablo Lemos,$^{1}$
William Hartley$^{1,2}$
and \newauthor
Jonathan Braden$^{3}$
\\
$^{1}$Department of Physics and Astronomy, University College London, Gower
Street, London WC1E 6BT, UK\\
$^{2}$Department of Physics, ETH Zurich, Wolfgang-Pauli-Strasse 16, CH-8093
Zurich, Switzerland\\
$^{3}$Canadian Institute for Theoretical Astrophysics, University of Toronto,
60 St. George St., Toronto, ON M5S 3H8, Canada
}

\date{Accepted XXX. Received YYY; in original form ZZZ}

\pubyear{2019}

\begin{document}
\label{firstpage}
\pagerange{\pageref{firstpage}--\pageref{lastpage}}
\maketitle

\begin{abstract}
In this work we investigate the systematic uncertainties that arise from the calculation of the peculiar velocity when estimating the Hubble constant ($H_0$) from gravitational wave standard sirens. We study the GW170817 event and the estimation of the peculiar velocity of its host galaxy, NGC 4993, when using Gaussian smoothing over nearby galaxies. NGC 4993 being a relatively nearby galaxy, at ${\sim}\,40~{\rm Mpc}$ away, is subject to a significant effect of peculiar velocities. We demonstrate a direct dependence of the estimated peculiar velocity value on the choice of smoothing scale. We show that when not accounting for this systematic, a bias of ${\sim}\,200~{\rm km\,s^{-1}}$ in the peculiar velocity incurs a bias of ${\sim}\,4~{\rm km\,s^{-1}\,Mpc^{-1}}$ on the Hubble constant. We formulate a Bayesian model that accounts for the dependence of the peculiar velocity on the smoothing scale and by marginalising over this parameter we remove the need for a choice of smoothing scale. The proposed model yields $H_0 = 68.6 ^{+14.0} _{-8.5}~{\rm km\,s^{-1}\,Mpc^{-1}}$. We demonstrate that under this model a more robust unbiased estimate of the Hubble constant from nearby GW sources is obtained.
\end{abstract}




\section{Introduction}
The present value of the Hubble constant ($H_0$) characterises the current expansion rate of the Universe. It is one of few cosmological parameters that can be estimated locally, and considerable resources have been dedicated to measuring its value with high precision. Despite this, its precise value is still a topic of controversy since the value inferred from measurements of the global Universe does not agree with the value obtained using local measurements. Using the cosmic microwave background (CMB), \citet{aghanim2018planck}  inferred $H_0 = 67.36 \pm 0.54$\,km\,s$^{-1}$\,Mpc$^{-1}$ providing the tightest constraint on the value of $H_0$ under the assumption of the standard $\Lambda$CDM cosmology. \citet{Riess2019}, using Cepheid variable stars and supernovae type Ia (SNIa) from Hubble Space Telescope (HST) observations, determined the local value of the Hubble constant to be $H_0 = 74.03 \pm 1.42$\,km\,s$^{-1}$\,Mpc$^{-1}$ providing a constraint independent of a cosmological model.\footnote{All quoted error bars represent the $68\%$ confidence level (CL), unless otherwise stated.} These latest results imply an increase in the discrepancy between the two precise methods to $4.4\,\sigma$. Whether this tension arises due to systematic effects or new physics is still debated \citep{Pourtsidou, Huang2016, Bernal_2016, valentino2016reconciling, Wyman, gomez2018h0}.  Another study using gravitationally lensed quasars with measured time delays (an independent method to SNIa and CMB) finds $H_0 = 73.3 ^{+1.7} _{-1.8}$\,km\,s$^{-1}$\,Mpc$^{-1}$ under the assumption of flat $\Lambda$CDM cosmology \citep{Holicow19}. While in agreement with \cite{Riess2019} it is in disagreement with the inferred value from CMB which further fuels this debate. Although significant effort has been made to reconcile the local and global estimates of $H_0$, no obvious systematic error accounting for the discrepancy has been reported \citep{efstathiou, wu2017sample, Follin, suhail, feeney2018clarifying}, and the underlying cause of the tension still remains elusive. 

Cosmology is in need of a new entirely independent method to effectively measure the Hubble constant. Gravitational waves (GWs) provide this, promising to offer key insight into this tension \citep{feeney2018prospects}. GWs can act as cosmological probes and carry enormous potential to examine the Universe and enhance our understanding of fundamental physical laws. The detection of GWs has been made possible with the advancement of the GW observatories Advanced LIGO (Laser Interferometer Gravitational-Wave Observatory) \citep{LIGO} and Advanced Virgo \citep{Virgo}. In the first and second observing runs, $11$ GW events were successfully observed \citep{LIGO_O1O2}. By comparing the detected GW signal to the waveforms predicted by general relativity, it is possible to extract the luminosity distance to the source \citep{lalinference}. Crucially, this does not rely on empirical relations used in conventional astronomical determinations of cosmological distances. The ability of GWs to act as distance indicators gave rise to the term standard sirens (analogous to SNIa standard candles) \citep{schutz1986}. In contrast to SNIa, which in conjunction with the distance ladder probe the distance-redshift relation, GWs do not uniquely provide a measure of the source's redshift due to the degeneracy between the rest-frame mass and the redshift, $z$, of the source. However, as first noted by \citet{schutz1986}, the most direct way to obtain the redshift of a GW source is through identifying an electromagnetic (EM) counterpart such as a glowing accretion disk or a gamma ray burst. In the case where a direct EM signal is not present, the redshift can still be obtained through a statistical approach by making use of galaxy catalogues to identify the potential host galaxies within the event localisation region \citep{schutz1986}. The redshifts of the potential host galaxies will then contribute in a probabilistic way to the calculation of $H_0$ \citep{delpozzo2012, chen2018two, fishbach2018standard, soares_firstH0}.

The first GW event to be accompanied by a direct EM counterpart was the binary neutron star (BNS) merger event, GW170817,  \citep{ligoGW170817observation, Abbott_2017_GR}. An intense follow-up observing campaign revealed an optical and near-infrared transient associated with the GW event, known as a kilonova, \citep{Abbott_2017_MM} which led to the identification of the host galaxy of the event, NGC 4993. The identification of the host galaxy contributed to the calculation of an independent estimate of the Hubble constant from this first BNS event of $H_0 = 70.0 ^{+12.0}_{-8.0}$\,km\,s$^{-1}$\,Mpc$^{-1}$ \citep[hereafter LVC17]{Abbott:2017nature}. While the measurement is broadly consistent with both the CMB and cosmic distance ladder results, this first multi-messenger event demonstrates the great potential of GW standard sirens to act as independent cosmological probes. Given LIGO's current $H_0$ estimate, a $2\%$ measurement of $H_0$ from standard sirens may be possible in the next ${\sim}\,5$ years, given ${\sim}\,50$ BNS merger events, sufficient to help clarify the current Hubble constant tension \citep{chen2018two, delpozzo2012, feeney2018prospects}. For GW cosmology to deliver on its promises, however, the $H_0$ estimate has to be unbiased,  free of systematics, and have representative uncertainties \citep{mortlock2018unbiased}. Taking these into account will lead to an era of precision GW cosmology.

The uncertainty in the Hubble constant from observations of nearby GW events (similar to GW170817) is dominated by the error on the peculiar velocity. In this paper we investigate the systematic uncertainties that arise from the calculation of the peculiar velocity and propose a way to limit their effect. This leads to a more robust calculation of the Hubble constant from nearby GW events. The remainder of this paper is organised as follows: the uncertainties associated with the Hubble constant calculation are discussed in Section~\ref{sec:uncertainties_in_H0}. In Section~\ref{sec:peculiar_velocity_calc} we demonstrate the existence of a previously unaccounted for systematic associated with the estimated peculiar velocity of the host galaxy from observations of its local neighbours from the 6dF Galaxy Survey \citep{springob20146df}. In Section~\ref{sec:bayesian_model} we introduce our new Bayesian model to control for this systematic, and compare it to the model used in LVC17 (referred to as baseline model). In Section~\ref{sec:results} we illustrate that, when using the baseline model, a bias of ${\sim}\,200$\,km\,s$^{-1}$ in the peculiar velocity incurs a bias of ${\sim}\,4$\,km\,s$^{-1}$\,Mpc$^{-1}$ on the Hubble constant making it impractical to clarify the $H_0$ tension. We demonstrate how this effect can be limited when using the proposed model instead. We then discuss and compare our results to other recent independent studies. Finally, our conclusions are summarised in Section~\ref{sec:concl} followed by Appendix~\ref{sec:pycbc} which outlines the settings used in {\tt PyCBC Inference} to infer the parameters of GW170817.

\section{Uncertainties in the Hubble Constant Calculation}
\label{sec:uncertainties_in_H0}

The Hubble flow velocity of an object is directly proportional to its distance and hence farther objects travel away from us at a greater velocity than nearby objects. This indicates that the Universe is expanding at a rate given by the Hubble constant \citep{hubble1929}. Galaxies, however, experience the local gravitational field which causes deviations in the galaxy's motion from the Hubble flow referred to as peculiar velocities. The peculiar velocity is defined as $v_p \equiv cz_p$, where $z_p$ is the peculiar velocity redshift which relates to the observed redshift $z_{obs}$ and the Hubble flow redshift $z_H$ through
\begin{equation}
    (1+z_{obs}) = (1+z_p)(1+z_H)\ .
    \label{eq:exact_redshifts}
\end{equation}

For small redshifts, $z << 1$, Equation~\ref{eq:exact_redshifts} approximates to 
\begin{equation}
    cz_{obs} \approx cz_H + cz_p \approx H_0 d + v_p\ ,
	\label{eq:hubblecorr}
\end{equation}
where $cz_{obs}=v_r$ is the observed recession velocity of the galaxy corresponding to observed redshift, $d$ is the distance to the object (which is distinct from the luminosity distance although equal in the $z<<1$ limit), and $v_p$ is the line-of-sight peculiar velocity. For nearby galaxies, the effect of the peculiar velocity is more significant compared to far away galaxies. This is because according to the Hubble law, more distant galaxies will have a larger Hubble flow velocity, which will be significantly greater than the induced peculiar motion. 

Rearranging Equation~\ref{eq:hubblecorr} we obtain
\begin{equation}
    H_0 \approx \frac{v_r-v_p}{d} = \frac{v}{d}\ ,
    \label{eq:hubblerearranged}
\end{equation}
where $v=v_r-v_p$ is the resultant velocity. The fractional error on the Hubble constant from a single source depends on the fractional uncertainty of the resultant velocity of the host galaxy and the fractional distance uncertainty to first order given by
\begin{equation}
    \left(\frac{\sigma_{H_0}}{H_0}\right)^2 \approx \left(\frac{\sigma_v}{v}\right)^2 + \left(\frac{\sigma_d}{d}\right)^2 .
    \label{eq:H0_uncertainty}
\end{equation}
$\sigma_{v}$ is dependent on the uncertainty of the recession velocity $\sigma_{v_r}$ and the uncertainty on the peculiar velocity $\sigma_{v_p}$ where typically $\sigma_{v_p} > \sigma_{v_r}$, hence the first term on the right hand sight of Equation~\ref{eq:H0_uncertainty} is dominated by the peculiar velocity uncertainty. As discussed in \citet{chen2018two}, at high redshifts $\sigma_{H_0}$ will be dominated by the distance uncertainty as $\sigma_{d}/d$ scales roughly inversely with signal-to-noise ratio (SNR) and hence tends to increase with distance. The distance uncertainty is expected to improve with next generation GW detectors and hence the point at which the two terms on the right hand side of Equation~\ref{eq:H0_uncertainty} are equal will shift to larger distances. For LIGO-Virgo's second observing run the distance at which the two terms were equal was ${\sim}\,30~{\rm Mpc}$ \citep{chen2018two}, close to the distance of GW170817 (${\sim}\,40$ Mpc). For sources found at such distances or closer, the uncertainty on the peculiar velocity (which can be as high as ${\sim}\,500$\,km\,s$^{-1}$) leads to poorer constraints on $H_0$ than for more distant events, despite typically having smaller localisation volumes \citep{ChenHolz16, Singer16_goingthedistance, LVC_localising_gws, palmese_astro2020}. While an increased distance reach will suppress the uncertainty on the peculiar velocity, as most events will come from farther away, it is possible that we are able to identify counterparts only for the closest GW events. Furthermore, observations of the superluminal motion of the counterpart jet from the Very Long Baseline Interferometer (VLBI) and the afterglow light curve data can substantially reduce the inclination angle-distance degeneracy as demonstrated by \cite{Hotokezaka19} making the peculiar velocity uncertainty more dominant. Therefore it is crucial to ensure the peculiar velocity calculation is unbiased, in order to achieve an unbiased estimate of the Hubble constant free of systematics.

\section{Peculiar Velocity of NGC 4993} 
\label{sec:peculiar_velocity_calc}
Using the EM counterpart signal, NGC 4993 was identified as the host galaxy of the BNS merger event, GW170817. In this section we inspect the calculation of the inferred value of the peculiar velocity of NGC 4993 from observations of its neighbours using the 6dF Galaxy survey \citep{springob20146df} and investigate the systematics that arise. The peculiar velocity analysis is carried out in the CMB frame. 

Calculating peculiar velocities is non-trivial. At present, there are two established methods for estimating peculiar velocities. The first method entails obtaining the redshift and distance measurements of galaxies and calculating the peculiar velocity directly via the use of Equation~\ref{eq:hubblecorr}. This method relies on the use of redshift-independent distance indicators such as the Tully-Fisher relation \citep{tullyfisher} and the fundamental plane relation \citep{FP_Dressler, FP_Djorgovski}. The former is a correlation that holds for spiral galaxies and expresses the luminosity of the galaxy as a power law function of its rotational velocity. The latter applies to galaxy spheroids and expresses a power law relationship between the effective radius of the galaxy, its surface brightness and its velocity dispersion. The second method for estimating peculiar velocities is by starting from a galaxy redshift survey and reconstructing the gravity vector  ${\bf g}$, at a position of interest by essentially summing up the inverse-square law over the catalogued galaxies. The peculiar velocity can be then estimated assuming  linear theory ${\bf v_p} \propto  (f/b) {\bf g}$ where $f$ is the growth factor and $b$ is the linear bias parameter, i.e. the ratio of density contrast in galaxies to mass \citep{Fisher1995, Davis1996, Erdogdu2006, springob20146df, carrick2015}.

\begin{figure}
	\includegraphics[width=\columnwidth]{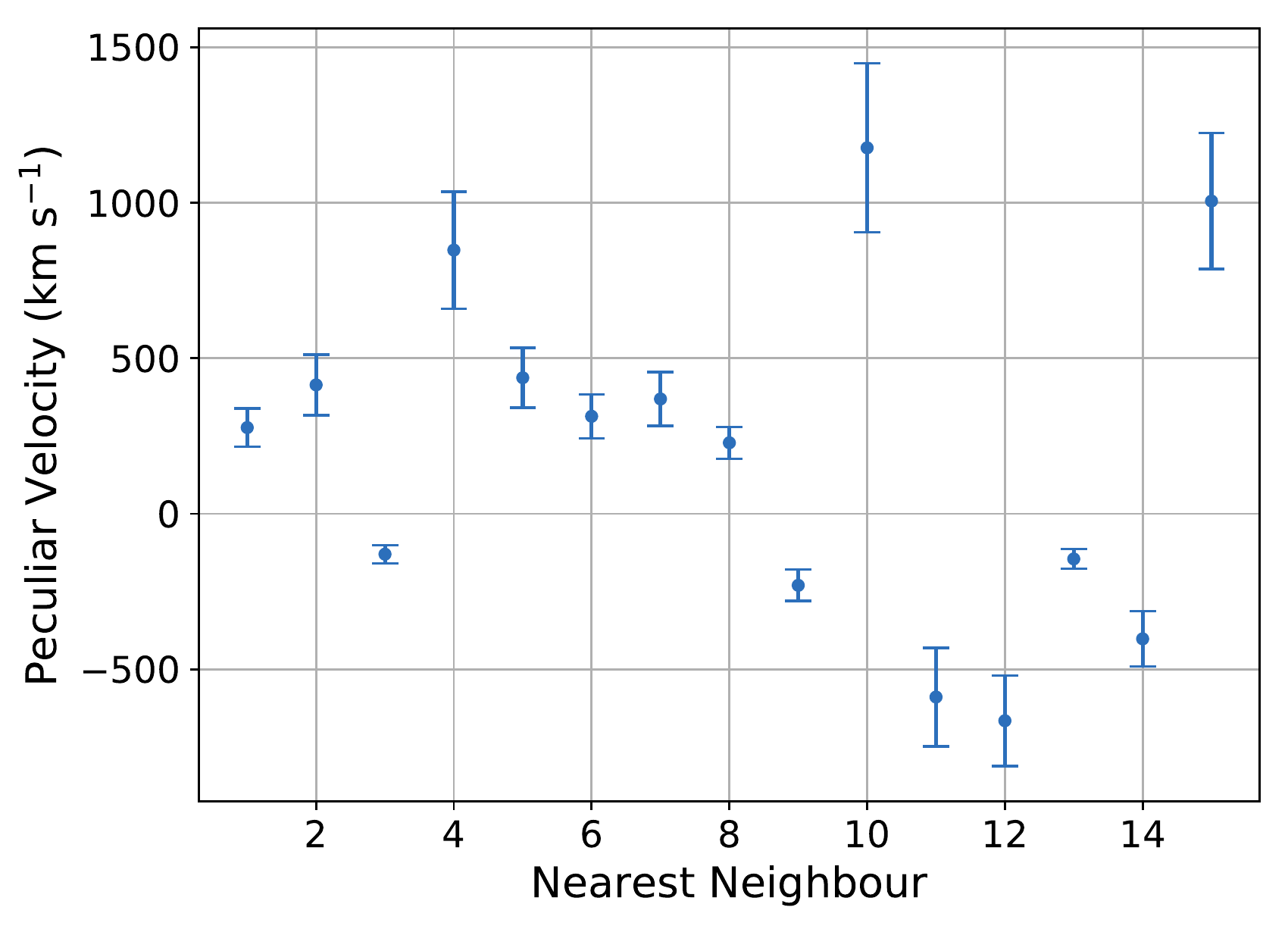}
    \includegraphics[width=\columnwidth]{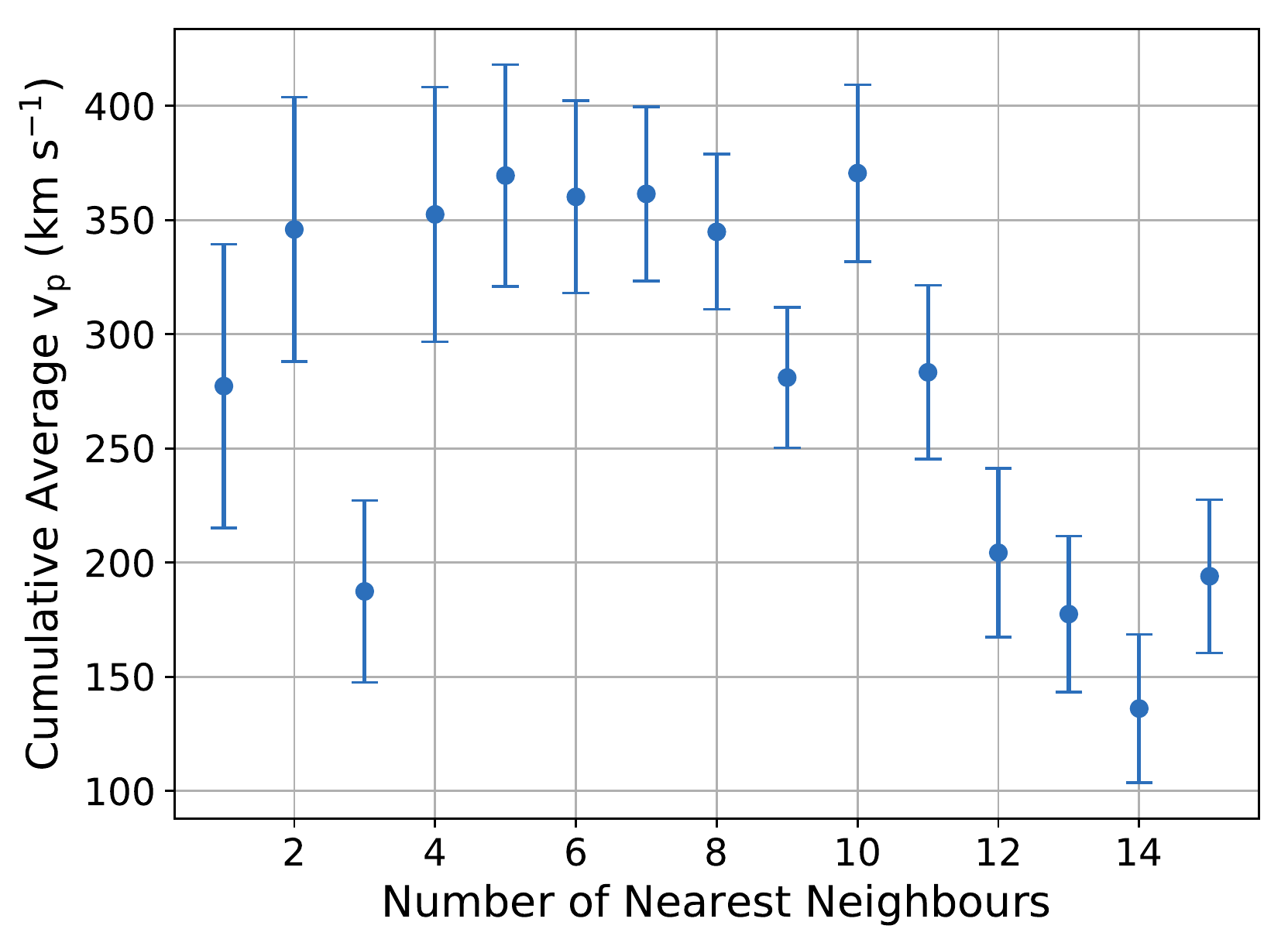}
    \caption{\textit{Top}: 15 nearest neighbours of NGC 4993 and their corresponding peculiar velocities with errors as obtained by the 6dFGSv. \textit{Bottom}: Cumulative average peculiar velocity as a function of the number of nearest neighbours. The error bars were obtained by error propagation from the individual peculiar velocity errors of the included galaxies. We note here that the error bars in both plots are an underestimate of the peculiar velocity error. A more representative error would be of the order of $150$\,km\,s$^{-1}$ or higher. The peculiar velocity analysis is carried out in the CMB frame.}
    \label{fig:nn_15}
\end{figure}

NGC 4993 appears to be part of a group of galaxies instead of being an isolated galaxy. There is evidence that it experienced a recent galaxy merger which \cite{Palmese17Formation} hypothesise led to the formation of the binary system, or inspiral of a pre-existing system, resulting from dynamical interactions. As the peculiar velocity of NGC 4993 is not directly available, a way of obtaining it is to infer the value from its neighbour galaxies. From the 3D peculiar velocity map of the 6dF Galaxy Survey (6dFGSv) \citep{springob20146df}, we can identify the nearest neighbours of NGC 4993. The top panel of Fig.~\ref{fig:nn_15} illustrates the 15 nearest galaxies and their corresponding radial peculiar velocity and the bottom panel the average radial peculiar velocity of NGC 4993 as a function of the number of nearest neighbours taken into account (equal weights are attributed to the galaxies in this case).

The top panel in Fig.~\ref{fig:nn_15} suggests that galaxies surrounding NGC 4993 are split into two categories: galaxies that possess a positive $v_p$ and galaxies that possess a negative $v_p$. This is reflected in the bottom panel of Fig.~\ref{fig:nn_15}, where the cumulative average peculiar velocity decreases when more than $10$ nearest neighbours are considered. Considering only the first $10$ nearest neighbours we obtain an average peculiar velocity of $371\pm39$\,km\,s$^{-1}$, whereas considering the nearest $15$ galaxies the average peculiar velocity is $194\pm34$\,km\,s$^{-1}$. This indicates that the average peculiar velocity obtained is highly dependent on the number of neighbours included, with relative variations on the order of ${\sim}\,50$\%. 

\begin{figure}
	\includegraphics[width=\columnwidth]{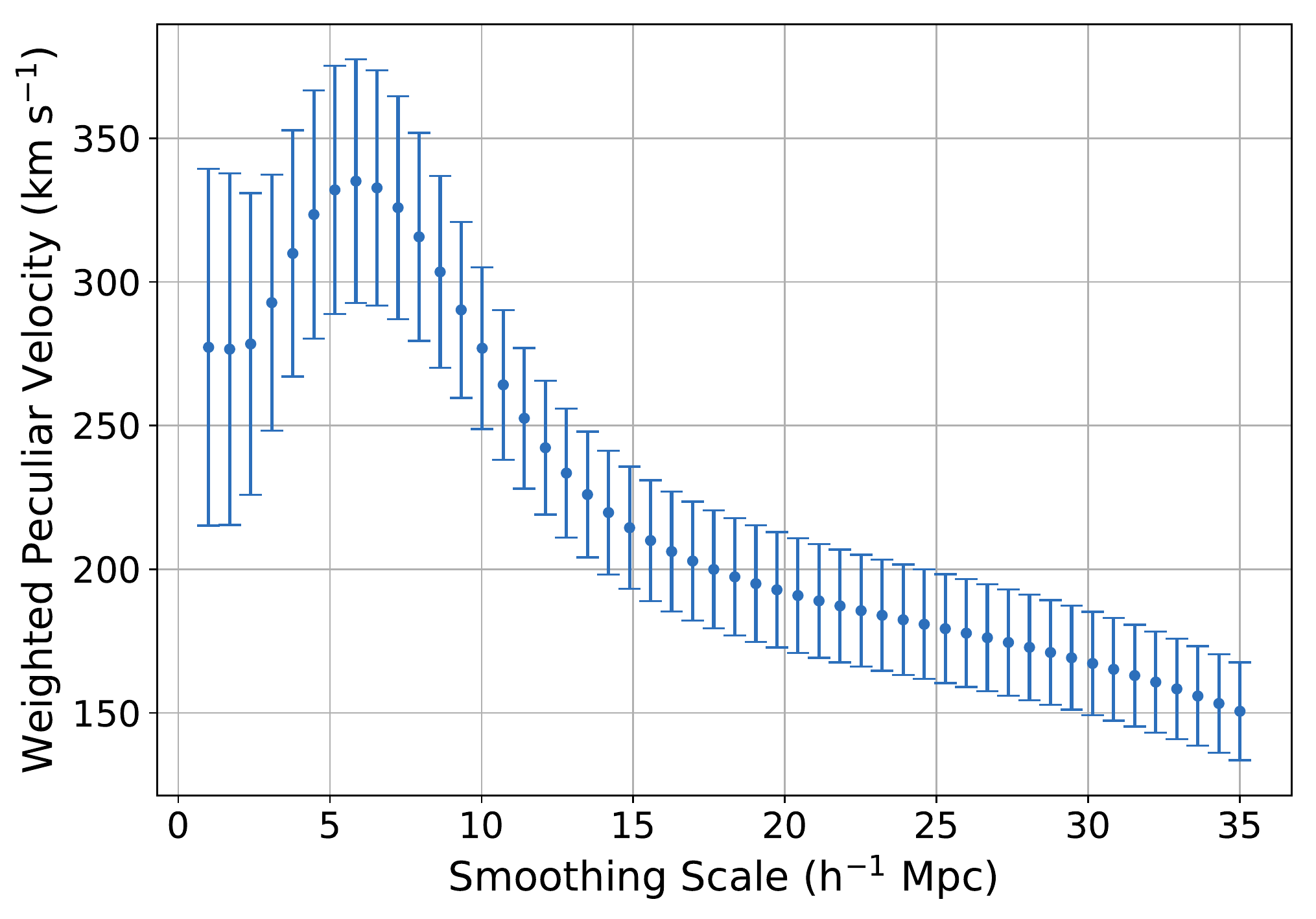}
    \caption{Weighted peculiar velocity as a function of the smoothing scale (width of  kernel) with $1\sigma$ error bars. Note that the 6dFGS error bars are underestimates. The points are correlated as galaxies that are included in smaller smoothing scales are also contributing in larger smoothing scales.}
    \label{fig:fwhm}
\end{figure}

Following LVC17, we adopt a 3D Gaussian kernel centred on the position of NGC 4993 to obtain the peculiar velocity of the host galaxy by weighing the peculiar velocities of the galaxies in the survey according to their distance from NGC 4993. Using a smoothing scale (width of kernel) of $8$\,h$^{-1}$\,Mpc, equivalent to $800$\,km\,s$^{-1}$ in velocity space, we obtain a weighted peculiar velocity of $315\pm36$\,km\,s$^{-1}$ (error bars indicate $1\sigma$). This is similar to the value obtained by LVC17 using the same smoothing scale ($310\pm69$\,km\,s$^{-1}$) with the difference being attributed to the fact that in this work we used the entire catalogue to obtain the weighted peculiar velocity whereas LVC17 use only the $10$ galaxies found within one kernel width of NGC 4993. As the choice of smoothing scale is arbitrary, we investigate how varying this parameter affects the calculation of the peculiar velocity. The results of this study are depicted in Fig.~\ref{fig:fwhm}.
We observe that the choice of smoothing scale affects the value of the resulting peculiar velocity substantially. This is expected as the smoothing scale determines the shape of the Gaussian kernel and consequently the distribution of weights. This suggests that the method incurs a systematic error as the choice of smoothing scale will bias the resulting peculiar velocity. The dependence depicted in Fig.~\ref{fig:fwhm} is not unique to NGC 4993. Inspecting other galaxies in the catalogue at random, we confirm a considerable dependence of the peculiar velocity on the smoothing scale with the exact mapping taking different forms depending on the chosen galaxy. In many cases this systematic affects the peculiar velocity estimate to a greater extent than shown in Fig.~\ref{fig:fwhm}.

We compare the above to other studies in the literature that independently obtain a peculiar velocity estimate for NGC 4993. \citet{guidorzi2017improved} estimate the peculiar velocity of NGC 4993 to be $326\pm250$\,km\,s$^{-1}$ where the peculiar velocity value was calculated using the 2MASS redshift survey \citep{carrick2015} and its error was computed via two methods \citep{wu2017sample, scolnic2018complete}. Two methods were employed because they suggest that the dispersion obtained from the 2MASS redshift survey is an underestimate because the method for obtaining the peculiar velocity is subject to systematics as it relies on the ability to convert from galaxy luminosity to the total matter field. It is interesting to note that the peculiar velocity value obtained by \citet{guidorzi2017improved} from the 2MASS redshift survey differs from the value obtained by LVC17 when using the 2MASS redshift survey ($280 \pm 150$\,km\,s$^{-1}$). The difference in the value is suggestive of a different smoothing scale used. \citet{hjorth2017distance} obtain an independent estimate of the peculiar velocity by making use of dark matter simulations from the Constrained Local Universe Simulations (CLUES) project. Using a $5$\,h$^{-1}$\,Mpc range centred at the CMB rest-frame velocity of NGC 4993 they find a mean peculiar velocity of $307\pm230$\,km\,s$^{-1}$. While all estimates of the peculiar velocity of NGC 4993 agree within $1\sigma$, the value is currently debated owing to a nontrivial dependence on the particular choice of reconstruction method, such as the arbitrary choice of the smoothing scale.

The results of the current work illustrate that the procedure to estimate the peculiar velocity of NGC 4993 is subject to a systematic uncertainty which, if left unaccounted for, will in turn bias the Hubble constant estimate. This systematic will apply to all methods that rely on the use of a smoothing scale to obtain the peculiar velocity. In the following section, we propose a method to limit this effect in order to obtain an unbiased estimate of the Hubble constant.

\section{Bayesian Model}
\label{sec:bayesian_model}
To obtain the posterior distribution of the Hubble constant, we construct a Bayesian model following LVC17 (referred to here as baseline model) and outline this work's proposed extension.

An observed GW event will generate a signal in the GW detectors, which we denote as $x_{GW}$. Suppose that we have also measured the recession velocity of the host (through an EM counterpart) and the mean peculiar velocity $\langle v_p\rangle$ of the neighbourhood of the host. As these observations are statistically independent the combined likelihood is

\begin{equation}
\begin{split}
	& p(x_{GW}, v_r, \langle v_p\rangle \mid d, \cos \iota, v_p, H_0) \\
	& = p(x_{GW} \mid d, \cos \iota) p(v_r \mid d, v_p, H_0)\,p(\langle v_p\rangle \mid v_p) .
	\label{eq:likelihood}
\end{split}
\end{equation}
The first term on the right hand side of Equation~\ref{eq:likelihood} is the parameter estimation likelihood of the observed GW data, marginalised over all parameters characterising the GW signal except $d$ and $\cos \iota$ (where $\iota$ is the inclination angle i.e. the angle between the binary's angular momentum axis and the line of sight). We compute this using {\tt PyCBC Inference}. See Appendix~\ref{sec:pycbc} for more details.

The quantity $p(v_r \mid d, v_p, H_0)$ is the likelihood of the recession velocity measurement and is modelled as

\begin{equation}
	p(v_r \mid d, v_p, H_0) = \mathcal{N}[H_0d+v_p,\,\sigma_{v_r}](v_r)\ ,
	\label{eq:vr_likelihood}
\end{equation}
where $\mathcal{N}[\mu,\,\sigma](x)$ is  a Gaussian probability density with mean $\mu$ and standard deviation $\sigma$, evaluated at $x$, i.e. the measured quantity. For the case of the host galaxy of GW170817, NGC 4993, we use the quoted value $v_r = 3327 \pm 72$\,km\,s$^{-1}$ from LVC17 \citep{Abbott:2017nature}.
A similar Gaussian likelihood is used for the measured peculiar
velocity

\begin{equation}
	p(\langle v_p\rangle \mid v_p) = \mathcal{N}[v_p,\,\sigma_{v_p}](\langle v_p\rangle) .
	\label{eq:vp_likelihood}
\end{equation}

From Equation~\ref{eq:likelihood} we derive the posterior

\begin{equation}
\begin{split}
	& p(H_0, d, \cos\iota, v_p \mid x_{GW}, v_r, \langle v_p\rangle) \propto p(x_{GW} \mid d, \cos \iota)\\
    & \times p(v_r \mid d, v_p, H_0)\,p(\langle v_p\rangle \mid v_p)\, p(H_0)\, p(d)\, p(\cos\iota)\, p(v_p)\ ,
	\label{eq:posterior}
\end{split}
\end{equation}
where $p(H_0)$, $p(d)$, $p(\cos\iota)$ and $p(v_p)$ are prior probabilities. GW analyses assume a uniform prior on the volume which translates to a prior on the distance $p(d) \propto d^2$. The domain of the distance prior is set here to be $[5, 80]$Mpc. We take a uniform prior on $\ln H_0$ i.e. $p(H_0) \propto 1/H_0$ (to allow for comparison with LVC17) in the range $[10,250]$\,km\,s$^{-1}$\,Mpc$^{-1}$, a uniform prior on $\cos\iota$, $p(\cos\iota) \propto \textrm{U}[-1, 1]$ and a uniform prior on $v_p$, $p(v_p) \propto \textrm{U}[-1000, 1000]$km\,s$^{-1}$ where $\textrm{U}[a, b]$ indicates a uniform distribution with lower bound $a$ and upper bound $b$. These priors characterise our beliefs regarding these parameters before any measurements and are chosen to be similar to the LVC17 analysis to allow for a comparison. A uniform prior on the peculiar velocity is suitable, as we don't have any information on its value prior to any measurements and hence assigning equal probabilities to a large range of possible velocities does not favour a particular value. Marginalising Equation~\ref{eq:posterior} over $v_p$, $d$ and $\cos\iota$ we obtain the marginalised posterior distribution of the Hubble constant

\begin{equation}
\begin{split}
	& p(H_0 \mid x_{GW}, v_r, \langle v_p\rangle) \propto p(H_0)\, \int p(x_{GW} \mid d, \cos \iota)\\
    & \times p(v_r \mid d, v_p, H_0)\,p(\langle v_p\rangle \mid v_p)\, p(d)\, p(\cos\iota)\, p(v_p)\\
    & \times \mathrm{d}v_p\, \mathrm{d}d\, \mathrm{d}\cos \iota\ . 
	\label{eq:h_posterior}
\end{split} 
\end{equation}

This baseline formalism is similar to the one used by LVC17. To capture the direct relation of the smoothing scale to the peculiar velocity we propose to introduce a parameter $s$, which represents the smoothing scale. This modifies the peculiar velocity likelihood given by Equation~\ref{eq:vp_likelihood} as follows

\begin{equation}
    p(\langle v_p\rangle \mid v_p, s) = \mathcal{N}[v_p,\,\sigma_{v_p}](\langle v_p\rangle(s)) \ ,
\end{equation}
Here, the chosen prior on the smoothing scale follows a Gamma distribution \citep{GammaWolfram} of shape $2$ and scale $4$\,h$^{-1}$\,Mpc i.e. $p(s) \propto \textrm{Gamma}[2, 4]$.
A shift of $1$\,h$^{-1}$\,Mpc is introduced as a scales between $0$ and $1$\,h$^{-1}$\,Mpc are unrepresentative. The smoothing scale prior has a maximum at $5$\,h$^{-1}$\,Mpc. This choice of prior represents typical non-linear smoothing scales. A prior that is too narrow may misrepresent the estimated peculiar velocity as it would penalise galaxies that belong to the same galaxy group but happen to be the members furthest from the galaxy in question. Similarly a prior looking at much larger scales may also disguise the true peculiar velocity value as it could amplify the effect of other galaxy groups further away. We tested the choice of prior and verified that the model is largely insensitive to prior specification (see Section~\ref{sec:results}).

The modified posterior distribution used in our improved model is
\begin{equation}
\begin{split}
    & p(H_0 \mid x_{GW}, v_r, \langle v_p\rangle) \propto p(H_0)\, \int p(x_{GW} \mid d, \cos \iota)\\
    & \times p(v_r \mid d, v_p, H_0)\,p(\langle v_p\rangle \mid v_p, s)\, p(s)\,p(d)\, p(\cos\iota)\\
    & \times p(v_p)\, \mathrm{d}v_p\, \mathrm{d}d\, \mathrm{d}\cos \iota\, \mathrm{d}s .
	\label{eq:mod_h_posterior}
\end{split}
\end{equation}

We utilise the {\tt emcee} Python package \citep{emcee} to perform the Markov Chain Monte Carlo (MCMC) sampling.

\section{Results and Discussion}
\label{sec:results}
\begin{figure}
	\includegraphics[width=\columnwidth]{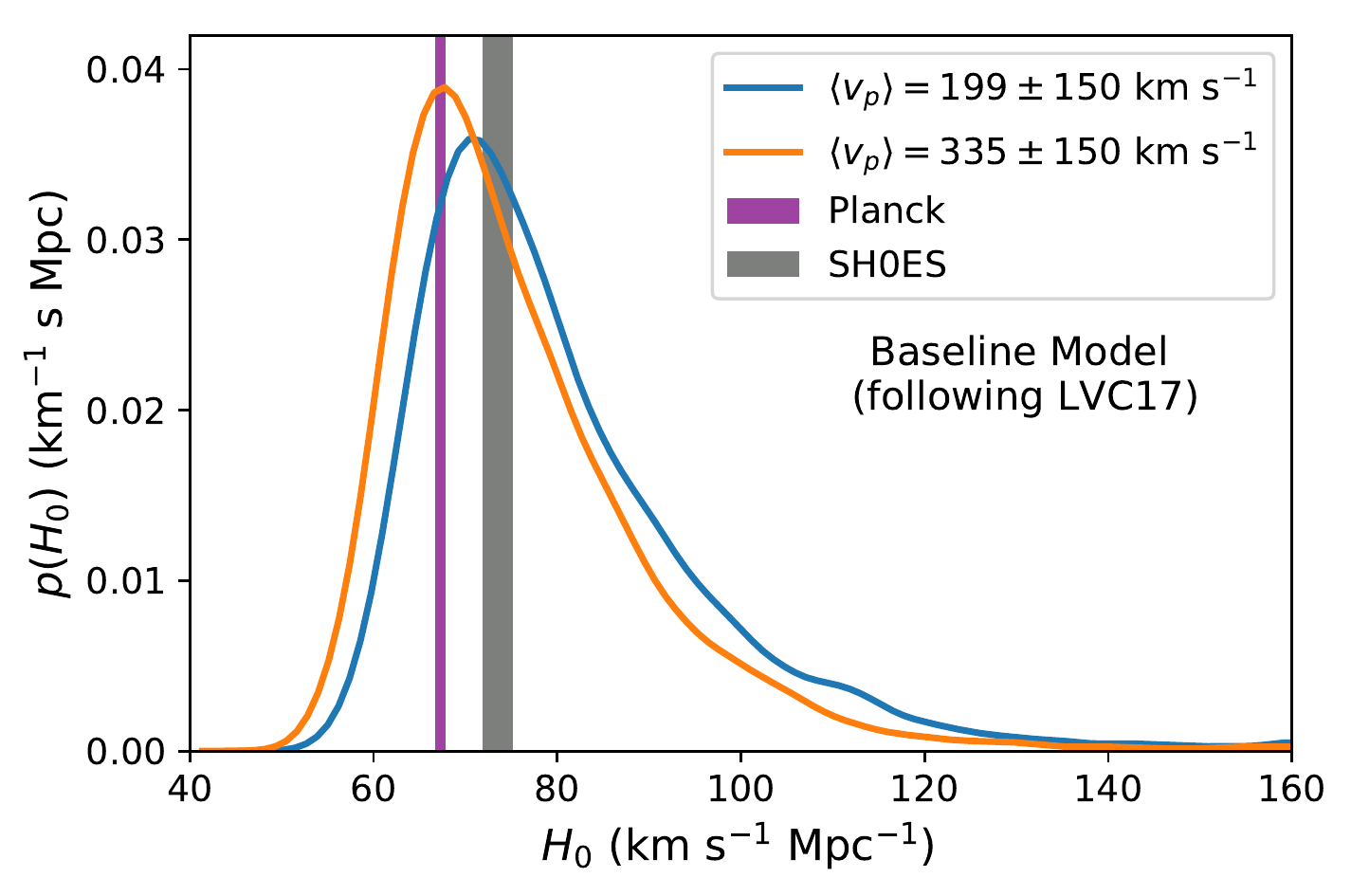}
	\includegraphics[width=\columnwidth]{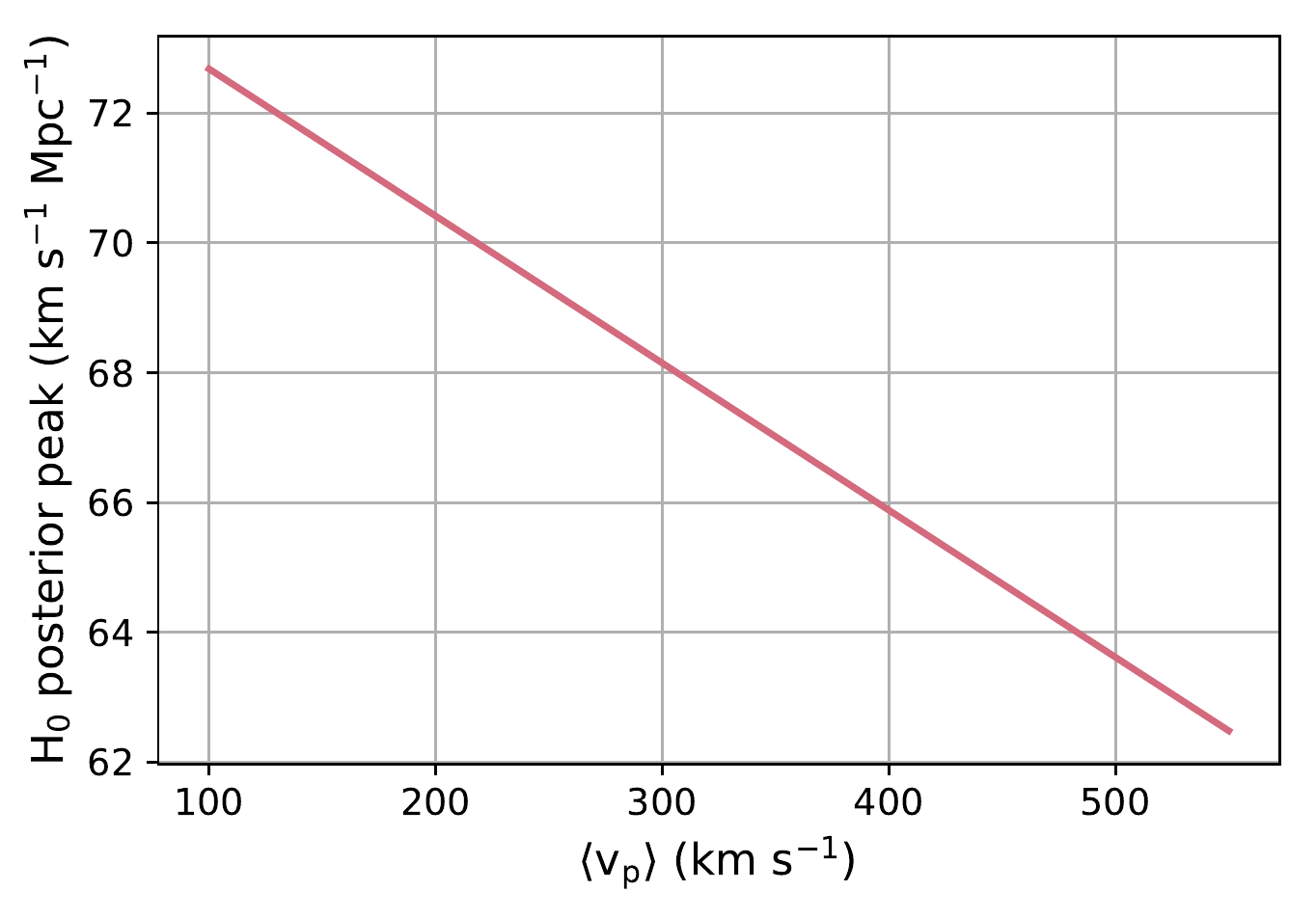}
    \caption{The top panel illustrates the $H_0$ posteriors computed using the baseline model for $\langle v_p \rangle = 199 \pm 150$\,km\,s$^{-1}$ obtained for $s = 6~{\rm h^{-1}\,Mpc}$ (blue line) and $\langle v_p \rangle = 335 \pm 150$\,km\,s$^{-1}$ for $s = 18~{\rm h^{-1}\,Mpc}$ (orange line) when using Gaussian smoothing from the 6dF galaxy survey. The vertical lines indicate the $1\sigma$ interval of the Planck (purple) and SH0ES (grey) results. The bottom panel illustrates the peak of the $H_0$ posterior distribution as a function of peculiar velocity under the baseline model.}
    \label{fig:baseline_model}
\end{figure}

To investigate the effect of the peculiar velocity value on the inferred value of the Hubble constant we use the baseline model as described in Section~\ref{sec:bayesian_model} for two different choices of smoothing scale: $6~{\rm h^{-1}\,Mpc}$ leading to $\langle v_p \rangle = 335 \pm 150~{\rm\,km\,s^{-1}}$ and $18~{\rm h^{-1}\,Mpc}$ leading to $\langle v_p \rangle = 199 \pm 150~{\rm\,km\,s^{-1}}$. Fig.~\ref{fig:baseline_model} (top panel) shows the $H_0$ posterior distribution for the two values of the peculiar velocity using the baseline model for the GW170817 event. For the case of $\langle v_p \rangle = 335 \pm 150~{\rm\,km\,s^{-1}}$ the maximum a posteriori value of the Hubble constant posterior along with the $68$\% HPD (Highest Posterior Density) interval is $H_0 = 67.4 ^{+13.7} _{-8.3}~{\rm km\,s^{-1}\,Mpc^{-1}}$ and for the case of $\langle v_p \rangle = 199 \pm 150~{\rm\,km\,s^{-1}}$, $H_0 = 70.7 ^{+15.4} _{-8.5}~{\rm km\,s^{-1}\,Mpc^{-1}}$. 

\begin{figure}
	\includegraphics[width=\columnwidth]{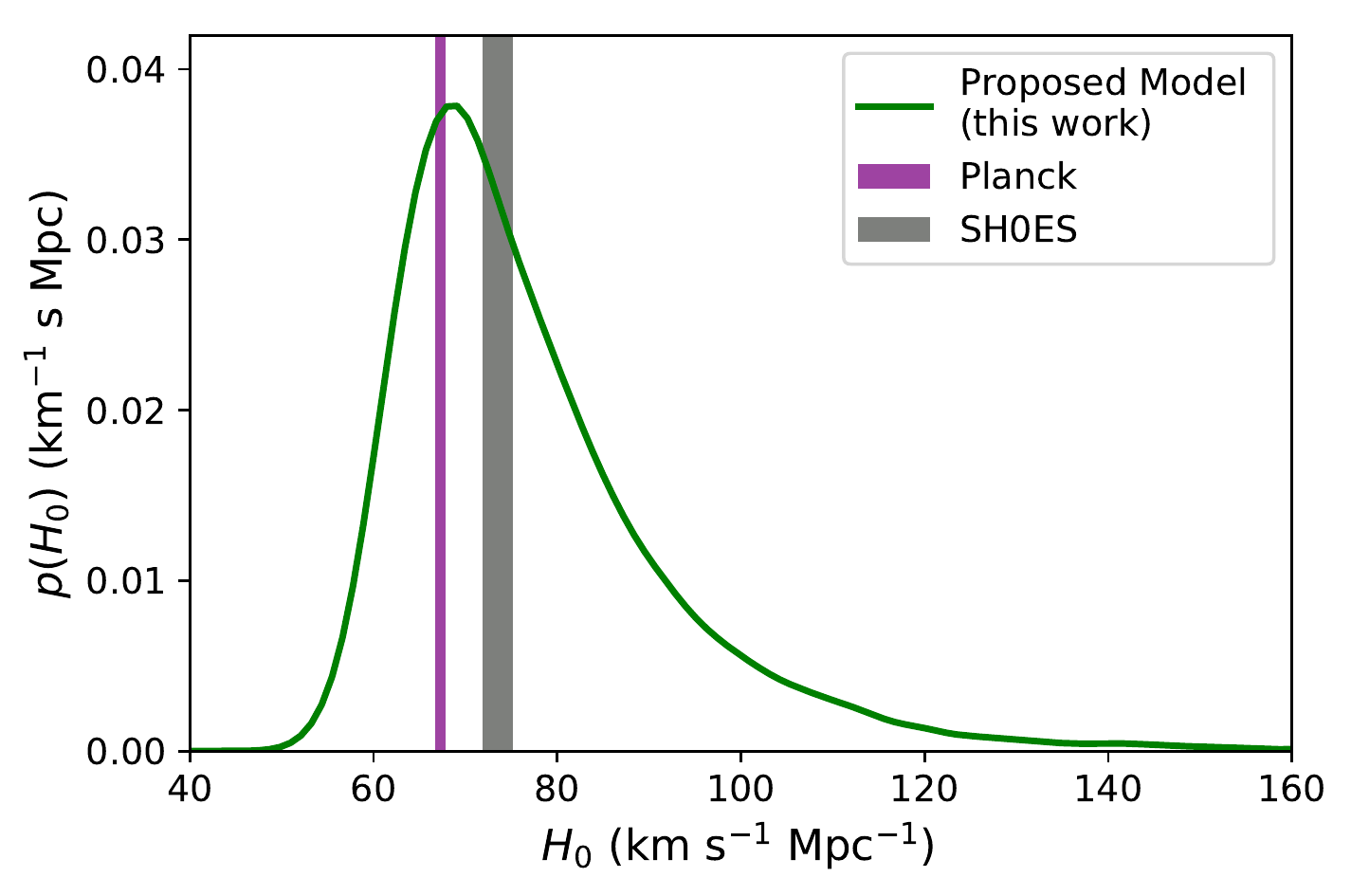}
    \caption{$H_0$ posterior obtained when using the proposed model. The vertical lines indicate the $1\sigma$ interval of the Planck (purple) and SH0ES (grey) results.}
    \label{fig:improved_model}
\end{figure}

We also plot the peak of the posterior distribution of $H_0$ as a function of peculiar velocity shown in the bottom panel of Fig.~\ref{fig:baseline_model}, recovering the linear relationship of Equation~\ref{eq:hubblerearranged}. It is evident that a bias of ${\sim}\,200$\,km\,s$^{-1}$ on the peculiar velocity due to a different choice of smoothing scale imparts a bias on the Hubble constant of ${\sim}\,4$\,km\,s$^{-1}$\,Mpc$^{-1}$, making it impractical to resolve the $H_0$ tension. This presents a systematic error in the current analysis of nearby GW sources to obtain $H_0$. While additional GW events promise to constrain the $H_0$ posterior to percent-level, this systematic will present a limitation.

To limit the effect of this systematic we propose the improved model described in Section~\ref{sec:bayesian_model}. Fig.~\ref{fig:improved_model} shows the posterior distribution of $H_0$ using the proposed model which yields $H_0 = 68.6 ^{+14.0} _{-8.5}~{\rm km\,s^{-1}\,Mpc^{-1}} $. The proposed model accounts for the relationship between the mean peculiar velocity and the smoothing scale, therefore removing the need to choose a specific smoothing scale in order to calculate the mean peculiar velocity. By incorporating the smoothing scale in the model we impose explicitly that the peculiar velocity is a function of the smoothing scale and hence the model yields a single posterior distribution accounting for the relation $\langle v_p \rangle (s)$. By imposing a prior on the smoothing scale which represents a reasonable range to be considered around the host galaxy and then marginalising over this parameter, the systematic that was present in the baseline model is no longer posing a limitation in the proposed model. Therefore, we obtain a more robust posterior distribution on the Hubble constant. It is also worth pointing out that the proposed model places similar constraints on $H_0$ as no significant increase in the error bars is observed.

We have shown that the proposed model provides a robust estimate of the Hubble constant. The applicability of this model is for the case where a direct measurement of the peculiar velocity of the host galaxy of a GW event is not available and hence the galaxy's peculiar velocity has to be inferred from neighbouring galaxies using Gaussian smoothing. However, in the case where a direct measurement of the peculiar velocity is available, it is important to obtain an accurate estimate, as a worse constraint on the peculiar velocity estimate directly translates to broader error bars on $H_0$. For the GW170817 event, if an accurate absolute measure of the peculiar velocity were available this would improve the uncertainty by ${\sim}\,2$\% of $H_0$. The improvement is only modest because the distance error dominates for this distance and the sensitivity of LIGO/Virgo when this event was observed (see Section~\ref{sec:uncertainties_in_H0}).

\subsection{Robustness to prior specification}
Our model uses a Gamma distribution with shape $2$ and scale $4$\,h$^{-1}$\,Mpc shifted by $1$\,h$^{-1}$\,Mpc for the smoothing scale prior. We investigate the effect of different choices of the prior on the posterior distribution of the Hubble constant to ensure robustness to prior specification. We test this using Gamma$[4, 2]$, Uniform$[1, 18]$ and Uniform$[1, 25]$. 

\begin{figure}
    \centering
    \includegraphics[width=\columnwidth]{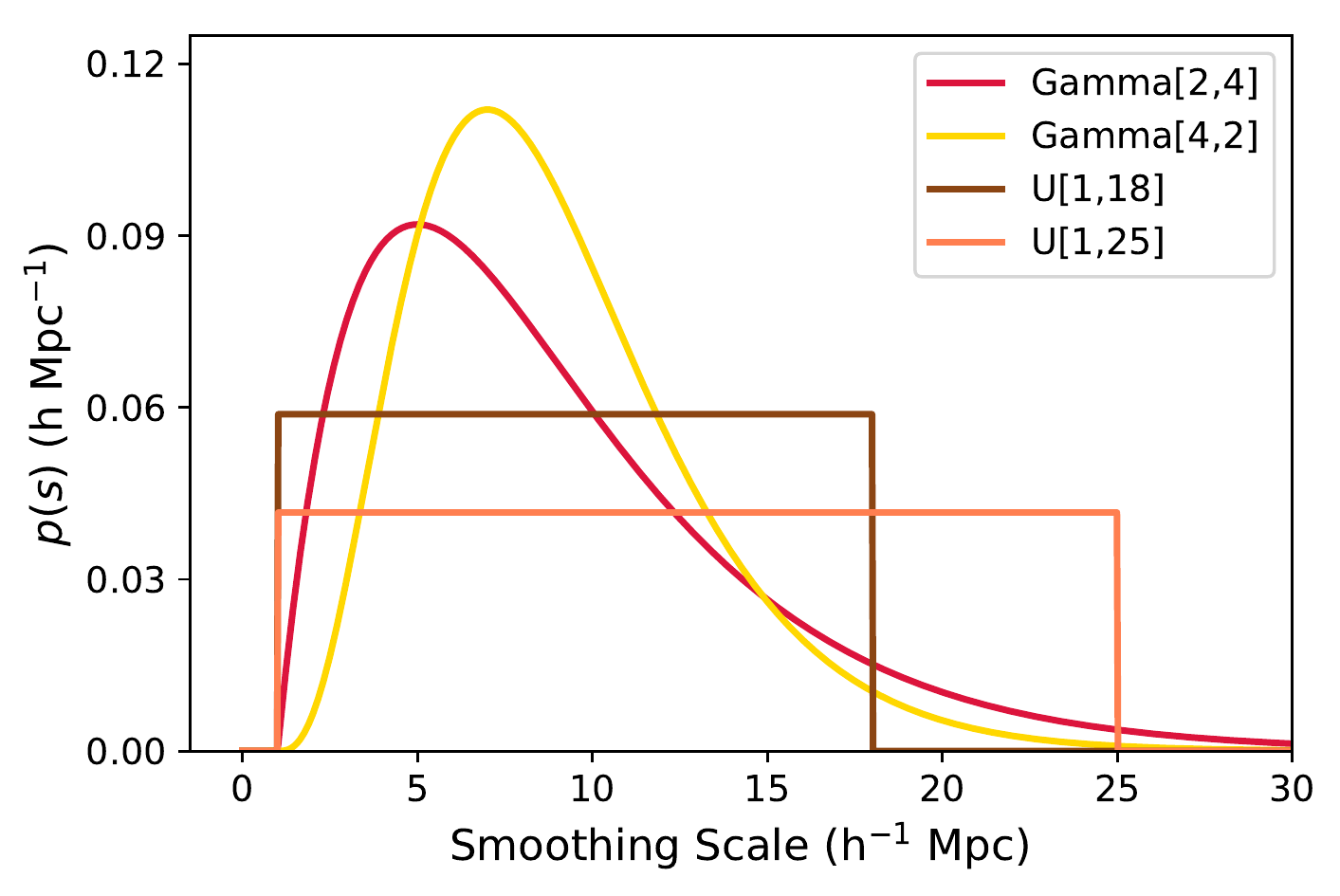}
    \includegraphics[width=\columnwidth]{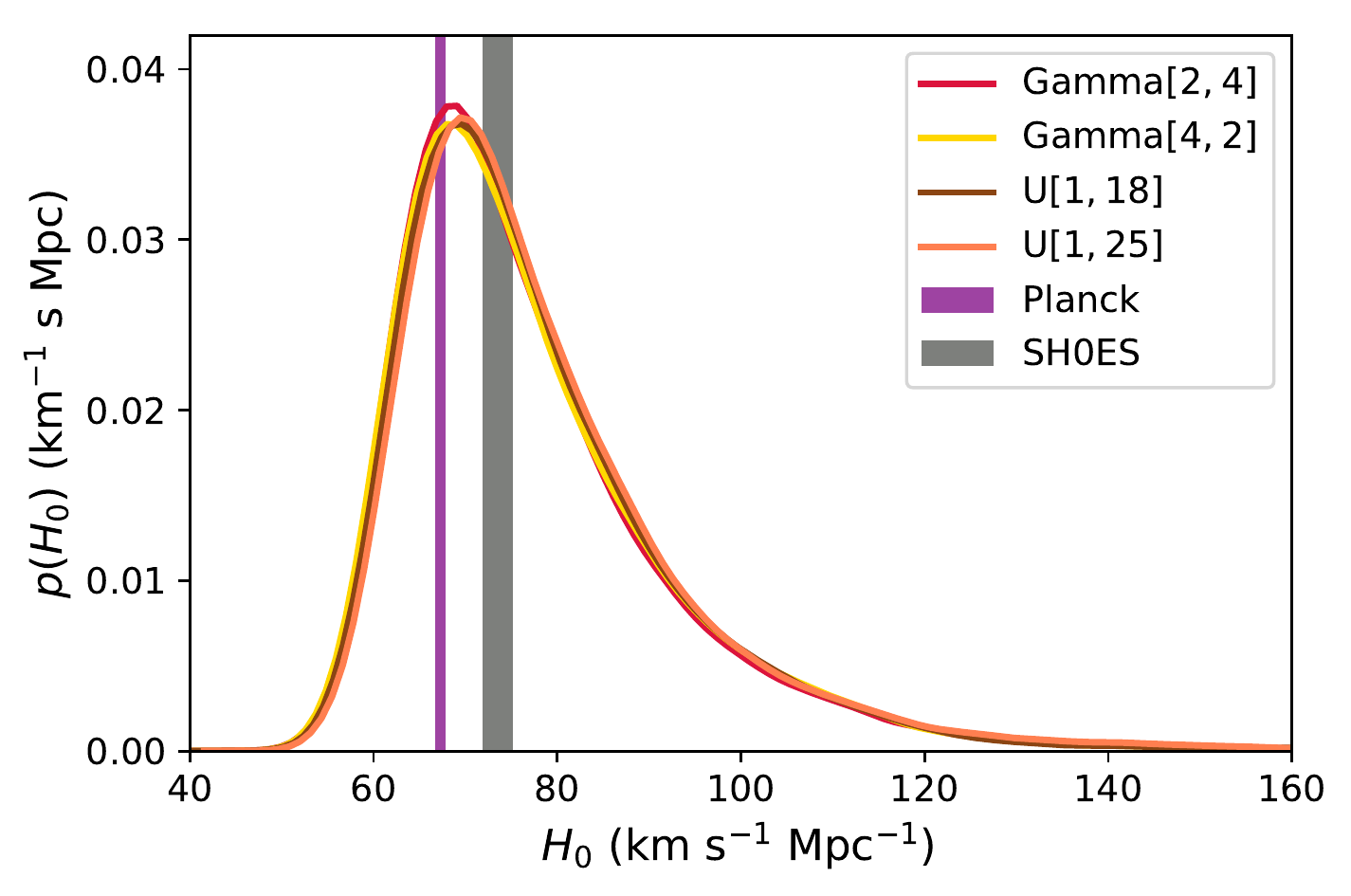}
    \caption{\textit{Top}: 4 different prior specifications for the smoothing scale: Gamma$[2, 4]$ (red), Gamma$[4, 2]$ (yellow), Uniform$[1,18]$ (brown) and Uniform$[1, 25]$ (orange). \textit{Bottom}: $H_0$ posterior distributions for the four different choices of prior specification on the smoothing scale. The vertical lines indicate the $1\sigma$ interval of the Planck (purple) and SH0ES (grey) results. }
    \label{fig:scale_prior}
\end{figure}

Fig.~\ref{fig:scale_prior} shows the prior distributions for the smoothing scale (top panel) and the corresponding $H_0$ posterior obtained from our Bayesian analysis (bottom panel). Since the four posterior distributions of $H_0$ are very similar, we conclude that the posterior distribution of $H_0$ is largely insensitive to the prior specification on the smoothing scale under our model.

\subsection{Comparison to recent independent studies}
As shown in this work, the arbitrary choice of smoothing scale leads to different values of the peculiar velocity for the host galaxy of the GW event, resulting in a bias in the $H_0$ posterior distribution. The concern raised and addressed in this work regarding the current methodology of calculating $H_0$ from GWs has also been raised by two other independent  studies \citep{Howlett19, Mukherjee19} that appeared almost concurrently with this paper strengthening the importance of correctly addressing the systematic error relating to the peculiar velocity value as shown in this paper, to achieve an unbiased estimate of the Hubble constant from GWs. Our methodology is unique as the other two papers take on slightly different approaches. Below we compare briefly our approach with these two papers.

\cite{Howlett19} refrain from using redshift approximations and instead use the actual redshift relations. This however imposes the assumption of a cosmological model; in their paper they assume a flat $\Lambda$CDM cosmology with $\Omega_m = 0.3$. Additionally in their model they work using the group velocities with the assumption that the size of the cluster is small enough that the distance to the centre of the cluster is approximately the distance to NGC 4993 itself. Also, their analysis is performed using the log-distance ratio $\eta = log_{10}(d(z_{cmb})/d({\bar{z}}))$ instead of using $v_p$. However, this quantity can only be used when the peculiar velocity estimate does not come from reconstructed fields. Repeating the LIGO analysis using their method and the same values as LVC, they obtain a small decrease of $< 0.5~{\rm km\,s^{-1}\,Mpc^{-1}} $ in the $H_0$ posterior which is not significant when compared to the uncertainty and the $68\%$ likelihood bounds. So using the accurate redshift formula, group properties and log-distance ratio instead of $v_p$ makes a negligible difference in the posterior of $H_0$.

The way \cite{Howlett19}  address the uncertainty in the peculiar and observed velocities is by identifying the possible observed velocities from 7 different group catalogues and the peculiar velocity from 3 galaxy catalogs (6dFGSv, 2MASS and CF3) as well as allowing the width of the smoothing kernel to vary between $2-8~{\rm h^{-1}\, Mpc}$ ending up with 154 models. Using Bayesian Model Averaging they perform a weighted average over all models using the Bayesian Evidence of each model to obtain a single posterior distribution on $H_0$. Their method yields $H_0 = 66.8 ^{+13.4} _{-9.2}~{\rm km\,s^{-1}\,Mpc^{-1}} $ which is ${\sim}\,3~{\rm km\,s^{-1}\,Mpc^{-1}} $ lower than the LVC17 result, largely due to the fact that a lot of the models included have a lower observed redshift and larger log-distance ratio than the canonical model. The main difference between the current work and \cite{Howlett19} is that we focus on the treatment of the smoothing scale as a consistent free parameter in our model, whereas \cite{Howlett19} include it as a set of discrete values specifying a collection of models which are then averaged over. The two papers complement each other in the sense that one could combine the two approaches by using our systematic marginalisation over the smoothing scale, reducing the computation time, with a Bayesian Average over the choice of group and peculiar velocity catalogues as outlined in \cite{Howlett19}.

\cite{Mukherjee19} approach the problem from a slightly different perspective. They propose to obtain the peculiar velocity of the host galaxy by using a statistical reconstruction method. They estimate the large scale velocity flow using BORG (Bayesian Origins Reconstruction from Galaxies) and the stochastic velocity dispersion using a numerical fitting-form which requires the mass of the halo of the host. Combining the two velocities gives the posterior distribution of the peculiar velocity. Their resulting peculiar velocity estimate $\langle v_p \rangle = 360 \pm 130~{\rm km\,s^{-1}}$ is $16\%$ higher and has a $13\%$ less standard deviation compared to LVC's $\langle v_p \rangle = 310 \pm 150~{\rm km\,s^{-1}}$. By combining their peculiar velocity estimate, their redshifts and the inferred luminosity distance from the GW data, they obtain a revised $H_0$ estimate where the posterior peak shifts slightly, to $69.3~{\rm km\,s^{-1}\,Mpc^{-1}} $ from LVC's $ H_0 = 70.0^{+12.0} _{-8.0}~{\rm km\,s^{-1}\,Mpc^{-1}}$ with the same $68.3\%$ credible interval obtained by LVC (see Figure 4 in \cite{Mukherjee19}).

The common ground between our paper and the two other independent papers, is that in order to achieve an unbiased estimate of the Hubble constant from nearby GWs we have to correctly account for the uncertainties in the estimation of the peculiar velocity of the host galaxy. Our methodology is unique as the other two papers take on slightly different approaches. The novel aspect of our approach is that we explicitly account for the relation between the peculiar velocity and the smoothing scale. This allows us to marginalise over the nuisance parameter, that is, the smoothing scale, while remaining cosmology independent.

\section{Conclusions}
\label{sec:concl}
Gravitational wave standard sirens are a new distance indicator offering the advantage of \textit{absolute} distance measures, unlike most extragalactic distance indicators. When used in conjunction with electromagnetic counterparts they can be used to independently determine the local value of the Hubble constant. Given the current tension on the value of $H_0$ between local and global estimates, GWs can offer key insight.

While future improvements of GW observatories will improve the distance estimate obtained from GWs leading to tighter constraints on the Hubble constant, nearby GWs will suffer from peculiar velocity uncertainties worsening the $H_0$ estimate. In this work, we studied the impact of possible systematic uncertainties in the calculation of the peculiar velocity of the host galaxy of the GW170817 merger event, NGC 4993. As a direct measurement of the peculiar velocity of NGC 4993 is not available, this was obtained by weighing the galaxies in the catalogue using Gaussian smoothing. We demonstrated the relationship between the smoothing scale and the resulting inferred peculiar velocity which induces a previously neglected systematic in the calculation of the peculiar velocity. When using the baseline model (following LVC17), a bias of ${\sim}\,200$\,km\,s$^{-1}$ in the peculiar velocity due to a different choice of smoothing scale incurs a bias of ${\sim}\,4$\,km\,s$^{-1}$\,Mpc$^{-1}$ on the Hubble constant making it impractical to help resolve the $H_0$ tension. This motivated us to introduce an improved model where the relationship between the smoothing scale and peculiar velocity is explicitly modelled. By doing so and marginalising over the smoothing scale, we obtain a more robust Hubble constant estimate, $H_0 = 68.6 ^{+14.0} _{-8.5}~{\rm km\,s^{-1}\,Mpc^{-1}} $, free of the arbitrary choice of smoothing scale. The coming years promise an abundance of GW events capable of constraining the Hubble constant to percent level accuracy. In the case where  a direct peculiar velocity measurement is not available, accounting for the systematic uncertainty induced by the choice of smoothing scale, as outlined in this work, is vital to ensuring the Hubble constant estimate from nearby standard sirens is unbiased.

\section*{Acknowledgements}
We thank Antonella Palmese and Yehuda Hoffman for helpful discussions. CN was supported by the STFC UCL Centre for Doctoral Training in Data Intensive Science (grant number ST/P006736/1). OL and PL acknowledge STFC Grant ST/R000476/1. JB is supported by the Simons Foundation Origins of the Universe program (Modern Inflationary Cosmology collaboration). JB was supported by the European Research Council (ERC) under the European Community's Seventh Framework Programme (FP7/2007-2013)/ERC
grant agreement number 3006478-CosmicDawn.



\bibliographystyle{mnras}
\bibliography{refs.bib}

\begin{thebibliography}{}
\makeatletter
\relax
\def\mn@urlcharsother{\let\do\@makeother \do\$\do\&\do\#\do\^\do\_\do\%\do\~}
\def\mn@doi{\begingroup\mn@urlcharsother \@ifnextchar [ {\mn@doi@}
  {\mn@doi@[]}}
\def\mn@doi@[#1]#2{\def\@tempa{#1}\ifx\@tempa\@empty \href
  {http://dx.doi.org/#2} {doi:#2}\else \href {http://dx.doi.org/#2} {#1}\fi
  \endgroup}
\def\mn@eprint#1#2{\mn@eprint@#1:#2::\@nil}
\def\mn@eprint@arXiv#1{\href {http://arxiv.org/abs/#1} {{\tt arXiv:#1}}}
\def\mn@eprint@dblp#1{\href {http://dblp.uni-trier.de/rec/bibtex/#1.xml}
  {dblp:#1}}
\def\mn@eprint@#1:#2:#3:#4\@nil{\def\@tempa {#1}\def\@tempb {#2}\def\@tempc
  {#3}\ifx \@tempc \@empty \let \@tempc \@tempb \let \@tempb \@tempa \fi \ifx
  \@tempb \@empty \def\@tempb {arXiv}\fi \@ifundefined
  {mn@eprint@\@tempb}{\@tempb:\@tempc}{\expandafter \expandafter \csname
  mn@eprint@\@tempb\endcsname \expandafter{\@tempc}}}

\bibitem[\protect\citeauthoryear{Abbott, Abbott, Abbott  et~al.}{Abbott
  et~al.}{2017a}]{ligoGW170817observation}
Abbott B.~P.,  Abbott R.,  Abbott T.~D.,   et~al., 2017a, \mn@doi [Phys. Rev.
  Lett.] {10.1103/PhysRevLett.119.161101}, 119, 161101

\bibitem[\protect\citeauthoryear{{Abbott} et~al.}{{Abbott}
  et~al.}{2017b}]{Abbott:2017nature}
{Abbott} B.~P.,  et~al., 2017b, \mn@doi [\nat] {10.1038/nature24471}, \href
  {https://ui.adsabs.harvard.edu/abs/2017Natur.551...85A} {551, 85}

\bibitem[\protect\citeauthoryear{Abbott et~al.}{Abbott
  et~al.}{2017c}]{Abbott_2017_MM}
Abbott B.~P.,  et~al., 2017c, \mn@doi [The Astrophysical Journal]
  {10.3847/2041-8213/aa91c9}, 848, L12

\bibitem[\protect\citeauthoryear{Abbott et~al.}{Abbott
  et~al.}{2017d}]{Abbott_2017_GR}
Abbott B.~P.,  et~al., 2017d, \mn@doi [The Astrophysical Journal]
  {10.3847/2041-8213/aa920c}, 848, L13

\bibitem[\protect\citeauthoryear{{Abbott} et~al.}{{Abbott}
  et~al.}{2018}]{LVC_localising_gws}
{Abbott} B.~P.,  et~al., 2018, \mn@doi [Living Reviews in Relativity]
  {10.1007/s41114-018-0012-9}, \href
  {https://ui.adsabs.harvard.edu/abs/2018LRR....21....3A} {21, 3}

\bibitem[\protect\citeauthoryear{Acernese et~al.}{Acernese
  et~al.}{2014}]{Virgo}
Acernese F.,  et~al., 2014, \mn@doi [Classical and Quantum Gravity]
  {10.1088/0264-9381/32/2/024001}, 32, 024001

\bibitem[\protect\citeauthoryear{Allen, Anderson, Brady, Brown  \&
  Creighton}{Allen et~al.}{2012}]{ALLEN_PSD_estimation}
Allen B.,  Anderson W.~G.,  Brady P.~R.,  Brown D.~A.,   Creighton J. D.~E.,
  2012, \mn@doi [Phys. Rev. D] {10.1103/PhysRevD.85.122006}, 85, 122006

\bibitem[\protect\citeauthoryear{Bernal, Verde  \& Riess}{Bernal
  et~al.}{2016}]{Bernal_2016}
Bernal J.~L.,  Verde L.,   Riess A.~G.,  2016, \mn@doi [Journal of Cosmology
  and Astroparticle Physics] {10.1088/1475-7516/2016/10/019}, 2016, 019

\bibitem[\protect\citeauthoryear{{Biwer}, {Capano}, {De}, {Cabero}, {Brown},
  {Nitz}  \& {Raymond}}{{Biwer} et~al.}{2019}]{biwer_pycbc}
{Biwer} C.~M.,  {Capano} C.~D.,  {De} S.,  {Cabero} M.,  {Brown} D.~A.,  {Nitz}
  A.~H.,   {Raymond} V.,  2019, \mn@doi [\pasp] {10.1088/1538-3873/aaef0b},
  \href {https://ui.adsabs.harvard.edu/abs/2019PASP..131b4503B} {131, 024503}

\bibitem[\protect\citeauthoryear{{Brown}, {Harry}, {Lundgren}  \&
  {Nitz}}{{Brown} et~al.}{2012}]{brown_bns_spin}
{Brown} D.~A.,  {Harry} I.,  {Lundgren} A.,   {Nitz} A.~H.,  2012, \mn@doi
  [\prd] {10.1103/PhysRevD.86.084017}, \href
  {https://ui.adsabs.harvard.edu/abs/2012PhRvD..86h4017B} {86, 084017}

\bibitem[\protect\citeauthoryear{{Buonanno}, {Iyer}, {Ochsner}, {Pan}  \&
  {Sathyaprakash}}{{Buonanno} et~al.}{2009}]{BuonannoTF2}
{Buonanno} A.,  {Iyer} B.~R.,  {Ochsner} E.,  {Pan} Y.,   {Sathyaprakash}
  B.~S.,  2009, \mn@doi [\prd] {10.1103/PhysRevD.80.084043}, \href
  {https://ui.adsabs.harvard.edu/abs/2009PhRvD..80h4043B} {80, 084043}

\bibitem[\protect\citeauthoryear{{Carrick}, {Turnbull}, {Lavaux}  \&
  {Hudson}}{{Carrick} et~al.}{2015}]{carrick2015}
{Carrick} J.,  {Turnbull} S.~J.,  {Lavaux} G.,   {Hudson} M.~J.,  2015, \mn@doi
  [\mnras] {10.1093/mnras/stv547}, \href
  {https://ui.adsabs.harvard.edu/abs/2015MNRAS.450..317C} {450, 317}

\bibitem[\protect\citeauthoryear{{Chen} \& {Holz}}{{Chen} \&
  {Holz}}{2016}]{ChenHolz16}
{Chen} H.-Y.,  {Holz} D.~E.,  2016, arXiv e-prints, \href
  {https://ui.adsabs.harvard.edu/abs/2016arXiv161201471C} {p. arXiv:1612.01471}

\bibitem[\protect\citeauthoryear{Chen, Fishbach  \& Holz}{Chen
  et~al.}{2018}]{chen2018two}
Chen H.-Y.,  Fishbach M.,   Holz D.~E.,  2018, \mn@doi [Nature]
  {10.1038/s41586-018-0606-0}, 562, 545

\bibitem[\protect\citeauthoryear{{Christensen} \& {Meyer}}{{Christensen} \&
  {Meyer}}{2001}]{christensen_meyer}
{Christensen} N.,  {Meyer} R.,  2001, \mn@doi [\prd]
  {10.1103/PhysRevD.64.022001}, \href
  {https://ui.adsabs.harvard.edu/abs/2001PhRvD..64b2001C} {64, 022001}

\bibitem[\protect\citeauthoryear{{Davis}, {Nusser}  \& {Willick}}{{Davis}
  et~al.}{1996}]{Davis1996}
{Davis} M.,  {Nusser} A.,   {Willick} J.~A.,  1996, \mn@doi [\apj]
  {10.1086/178124}, \href
  {https://ui.adsabs.harvard.edu/abs/1996ApJ...473...22D} {473, 22}

\bibitem[\protect\citeauthoryear{Del~Pozzo}{Del~Pozzo}{2012}]{delpozzo2012}
Del~Pozzo W.,  2012, \mn@doi [Phys. Rev. D] {10.1103/PhysRevD.86.043011}, 86,
  043011

\bibitem[\protect\citeauthoryear{{Dhawan, Suhail}, {Jha, Saurabh W.}  \&
  {Leibundgut, Bruno}}{{Dhawan, Suhail} et~al.}{2018}]{suhail}
{Dhawan, Suhail} {Jha, Saurabh W.}  {Leibundgut, Bruno} 2018, \mn@doi [A\&A]
  {10.1051/0004-6361/201731501}, 609, A72

\bibitem[\protect\citeauthoryear{Di~Valentino, Melchiorri  \&
  Silk}{Di~Valentino et~al.}{2016}]{valentino2016reconciling}
Di~Valentino E.,  Melchiorri A.,   Silk J.,  2016, \mn@doi [Phys. Lett. B]
  {10.1016/j.physletb.2016.08.043}, 761, 242

\bibitem[\protect\citeauthoryear{{Djorgovski} \& {Davis}}{{Djorgovski} \&
  {Davis}}{1987}]{FP_Djorgovski}
{Djorgovski} S.,  {Davis} M.,  1987, \mn@doi [\apj] {10.1086/164948}, \href
  {https://ui.adsabs.harvard.edu/abs/1987ApJ...313...59D} {313, 59}

\bibitem[\protect\citeauthoryear{{Dressler}, {Lynden-Bell}, {Burstein},
  {Davies}, {Faber}, {Terlevich}  \& {Wegner}}{{Dressler}
  et~al.}{1987}]{FP_Dressler}
{Dressler} A.,  {Lynden-Bell} D.,  {Burstein} D.,  {Davies} R.~L.,  {Faber}
  S.~M.,  {Terlevich} R.,   {Wegner} G.,  1987, \mn@doi [\apj]
  {10.1086/164947}, \href
  {https://ui.adsabs.harvard.edu/abs/1987ApJ...313...42D} {313, 42}

\bibitem[\protect\citeauthoryear{Efstathiou}{Efstathiou}{2014}]{efstathiou}
Efstathiou G.,  2014, \mn@doi [Monthly Notices of the Royal Astronomical
  Society] {10.1093/mnras/stu278}, 440, 1138

\bibitem[\protect\citeauthoryear{Erdogdu et~al.,}{Erdogdu
  et~al.}{2006}]{Erdogdu2006}
Erdogdu P.,  et~al., 2006, \mn@doi [Monthly Notices of the Royal Astronomical
  Society] {10.1111/j.1365-2966.2006.11049.x}, 373, 45

\bibitem[\protect\citeauthoryear{Feeney, Peiris, Williamson, Nissanke,
  Mortlock, Alsing  \& Scolnic}{Feeney et~al.}{2018a}]{feeney2018prospects}
Feeney S.~M.,  Peiris H.~V.,  Williamson A.~R.,  Nissanke S.~M.,  Mortlock
  D.~J.,  Alsing J.,   Scolnic D.,  2018a, arXiv:1802.03404 [astro-ph.CO]

\bibitem[\protect\citeauthoryear{Feeney, Mortlock  \& Dalmasso}{Feeney
  et~al.}{2018b}]{feeney2018clarifying}
Feeney S.~M.,  Mortlock D.~J.,   Dalmasso N.,  2018b, \mn@doi [MNRAS]
  {10.1093/mnras/sty418}, 476, 3861

\bibitem[\protect\citeauthoryear{Fishbach, Gray, Hernandez, Qi, Sur
  et~al.}{Fishbach et~al.}{2018}]{fishbach2018standard}
Fishbach M.,  Gray R.,  Hernandez I.~M.,  Qi H.,  Sur A.,   et~al., 2018, arXiv
  preprint arXiv:1807.05667

\bibitem[\protect\citeauthoryear{Fisher, Lahav, Hoffman, Lynden-Bell  \&
  Zaroubi}{Fisher et~al.}{1995}]{Fisher1995}
Fisher K.~B.,  Lahav O.,  Hoffman Y.,  Lynden-Bell D.,   Zaroubi S.,  1995,
  \mn@doi [Monthly Notices of the Royal Astronomical Society]
  {10.1093/mnras/272.4.885}, 272, 885

\bibitem[\protect\citeauthoryear{Follin \& Knox}{Follin \& Knox}{2018}]{Follin}
Follin B.,  Knox L.,  2018, \mn@doi [Monthly Notices of the Royal Astronomical
  Society] {10.1093/mnras/sty720}, 477, 4534

\bibitem[\protect\citeauthoryear{{Foreman-Mackey}, {Hogg}, {Lang}  \&
  {Goodman}}{{Foreman-Mackey} et~al.}{2013}]{emcee}
{Foreman-Mackey} D.,  {Hogg} D.~W.,  {Lang} D.,   {Goodman} J.,  2013, \mn@doi
  [PASP] {10.1086/670067}, 125, 306

\bibitem[\protect\citeauthoryear{G{\'o}mez-Valent \& Amendola}{G{\'o}mez-Valent
  \& Amendola}{2018}]{gomez2018h0}
G{\'o}mez-Valent A.,  Amendola L.,  2018, \mn@doi [JCAP]
  {10.1088/1475-7516/2018/04/051}, 2018, 051

\bibitem[\protect\citeauthoryear{Guidorzi et~al.,}{Guidorzi
  et~al.}{2017}]{guidorzi2017improved}
Guidorzi C.,  et~al., 2017, \mn@doi [ApJL] {10.3847/2041-8213/aaa009}, 851, L36

\bibitem[\protect\citeauthoryear{Hjorth et~al.,}{Hjorth
  et~al.}{2017}]{hjorth2017distance}
Hjorth J.,  et~al., 2017, \mn@doi [ApJL] {10.3847/2041-8213/aa9110}, 848, L31

\bibitem[\protect\citeauthoryear{{Hotokezaka}, {Nakar}, {Gottlieb}, {Nissanke},
  {Masuda}, {Hallinan}, {Mooley}  \& {Deller}}{{Hotokezaka}
  et~al.}{2019}]{Hotokezaka19}
{Hotokezaka} K.,  {Nakar} E.,  {Gottlieb} O.,  {Nissanke} S.,  {Masuda} K.,
  {Hallinan} G.,  {Mooley} K.~P.,   {Deller} A.~T.,  2019, \mn@doi [Nature
  Astronomy] {10.1038/s41550-019-0820-1}, \href
  {https://ui.adsabs.harvard.edu/abs/2019NatAs...3..940H} {3, 940}

\bibitem[\protect\citeauthoryear{Howlett \& Davis}{Howlett \&
  Davis}{2020}]{Howlett19}
Howlett C.,  Davis T.~M.,  2020, \mn@doi [Monthly Notices of the Royal
  Astronomical Society] {10.1093/mnras/staa049}, 492, 3803

\bibitem[\protect\citeauthoryear{Huang \& Wang}{Huang \&
  Wang}{2016}]{Huang2016}
Huang Q.-G.,  Wang K.,  2016, \mn@doi [The European Physical Journal C]
  {10.1140/epjc/s10052-016-4352-x}, 76, 506

\bibitem[\protect\citeauthoryear{Hubble}{Hubble}{1929}]{hubble1929}
Hubble E.,  1929, \mn@doi [PNAS] {10.1073/pnas.15.3.168}, 15, 168

\bibitem[\protect\citeauthoryear{{LIGO Scientific Collaboration} et~al.,}{{LIGO
  Scientific Collaboration} et~al.}{2015}]{LIGO}
{LIGO Scientific Collaboration} et~al., 2015, \mn@doi [Classical and Quantum
  Gravity] {10.1088/0264-9381/32/7/074001}, \href
  {https://ui.adsabs.harvard.edu/abs/2015CQGra..32g4001L} {32, 074001}

\bibitem[\protect\citeauthoryear{{Mortlock}, {Feeney}, {Peiris}, {Williamson}
  \& {Nissanke}}{{Mortlock} et~al.}{2018}]{mortlock2018unbiased}
{Mortlock} D.~J.,  {Feeney} S.~M.,  {Peiris} H.~V.,  {Williamson} A.~R.,
  {Nissanke} S.~M.,  2018, arXiv e-prints, \href
  {https://ui.adsabs.harvard.edu/abs/2018arXiv181111723M} {p. arXiv:1811.11723}

\bibitem[\protect\citeauthoryear{Mukherjee, Lavaux, Bouchet, Jasche, Wandelt,
  Nissanke, Leclercq  \& Hotokezaka}{Mukherjee et~al.}{2019}]{Mukherjee19}
Mukherjee S.,  Lavaux G.,  Bouchet F.~R.,  Jasche J.,  Wandelt B.~D.,  Nissanke
  S.~M.,  Leclercq F.,   Hotokezaka K.,  2019

\bibitem[\protect\citeauthoryear{Nitz et~al.,}{Nitz et~al.}{2018}]{nitz_pycbc}
Nitz A.,  et~al., 2018, ligo-cbc/pycbc: Post-O2 Release 6,
  \mn@doi{10.5281/zenodo.1183449}

\bibitem[\protect\citeauthoryear{{Palmese} et~al.,}{{Palmese}
  et~al.}{2017}]{Palmese17Formation}
{Palmese} A.,  et~al., 2017, \mn@doi [\apjl] {10.3847/2041-8213/aa9660}, \href
  {https://ui.adsabs.harvard.edu/abs/2017ApJ...849L..34P} {849, L34}

\bibitem[\protect\citeauthoryear{{Palmese} et~al.,}{{Palmese}
  et~al.}{2019}]{palmese_astro2020}
{Palmese} A.,  et~al., 2019, arXiv e-prints, \href
  {https://ui.adsabs.harvard.edu/abs/2019arXiv190304730P} {p. arXiv:1903.04730}

\bibitem[\protect\citeauthoryear{{Planck Collaboration}}{{Planck
  Collaboration}}{2018}]{aghanim2018planck}
{Planck Collaboration} 2018, arXiv e-prints, \href
  {https://ui.adsabs.harvard.edu/abs/2018arXiv180706209P} {p. arXiv:1807.06209}

\bibitem[\protect\citeauthoryear{Pourtsidou \& Tram}{Pourtsidou \&
  Tram}{2016}]{Pourtsidou}
Pourtsidou A.,  Tram T.,  2016, \mn@doi [Phys. Rev. D]
  {10.1103/PhysRevD.94.043518}, 94, 043518

\bibitem[\protect\citeauthoryear{{Riess}, {Casertano}, {Yuan}, {Macri}  \&
  {Scolnic}}{{Riess} et~al.}{2019}]{Riess2019}
{Riess} A.~G.,  {Casertano} S.,  {Yuan} W.,  {Macri} L.~M.,   {Scolnic} D.,
  2019, \mn@doi [\apj] {10.3847/1538-4357/ab1422}, \href
  {https://ui.adsabs.harvard.edu/abs/2019ApJ...876...85R} {876, 85}

\bibitem[\protect\citeauthoryear{Schutz}{Schutz}{1986}]{schutz1986}
Schutz B.~F.,  1986, \mn@doi [Nature] {10.1038/323310a0}, 323, 310

\bibitem[\protect\citeauthoryear{Scolnic et~al.,}{Scolnic
  et~al.}{2018}]{scolnic2018complete}
Scolnic D.,  et~al., 2018, \mn@doi [ApJ] {10.3847/1538-4357/aab9bb}, 859, 101

\bibitem[\protect\citeauthoryear{{Singer} et~al.,}{{Singer}
  et~al.}{2016}]{Singer16_goingthedistance}
{Singer} L.~P.,  et~al., 2016, \mn@doi [\apjl] {10.3847/2041-8205/829/1/L15},
  \href {https://ui.adsabs.harvard.edu/abs/2016ApJ...829L..15S} {829, L15}

\bibitem[\protect\citeauthoryear{{{Soares-Santos} \& {Palmese}}
  et~al.,}{{{Soares-Santos} \& {Palmese}} et~al.}{2019}]{soares_firstH0}
{{Soares-Santos} \& {Palmese}} et~al., 2019, arXiv e-prints, \href
  {https://ui.adsabs.harvard.edu/abs/2019arXiv190101540T} {p. arXiv:1901.01540}

\bibitem[\protect\citeauthoryear{Soares-Santos et~al.,}{Soares-Santos
  et~al.}{2017}]{soares2017}
Soares-Santos M.,  et~al., 2017, \mn@doi [ApJL] {10.3847/2041-8213/aa9059},
  848, L16

\bibitem[\protect\citeauthoryear{Springob et~al.,}{Springob
  et~al.}{2014}]{springob20146df}
Springob C.~M.,  et~al., 2014, \mn@doi [MNRAS] {10.1093/mnras/stu1743}, 445,
  2677

\bibitem[\protect\citeauthoryear{{The LIGO Scientific Collaboration}, {the
  Virgo Collaboration}, {Abbott}, {Abbott}, {Abbott}  et~al.}{{The LIGO
  Scientific Collaboration} et~al.}{2018}]{LIGO_O1O2}
{The LIGO Scientific Collaboration} {the Virgo Collaboration} {Abbott} B.~P.,
  {Abbott} R.,  {Abbott} T.~D.,   et~al., 2018, arXiv e-prints, \href
  {https://ui.adsabs.harvard.edu/abs/2018arXiv181112907T} {p. arXiv:1811.12907}

\bibitem[\protect\citeauthoryear{{Tully} \& {Fisher}}{{Tully} \&
  {Fisher}}{1977}]{tullyfisher}
{Tully} R.~B.,  {Fisher} J.~R.,  1977, \aap, \href
  {https://ui.adsabs.harvard.edu/abs/1977A%26A....54..661T} {54, 661}

\bibitem[\protect\citeauthoryear{Vallisneri, Kanner, Williams, Weinstein  \&
  Stephens}{Vallisneri et~al.}{2015}]{Vallisneri_losc}
Vallisneri M.,  Kanner J.,  Williams R.,  Weinstein A.,   Stephens B.,  2015,
  \mn@doi [Journal of Physics: Conference Series]
  {10.1088/1742-6596/610/1/012021}, 610, 012021

\bibitem[\protect\citeauthoryear{Veitch et~al.,}{Veitch
  et~al.}{2015}]{lalinference}
Veitch J.,  et~al., 2015, \mn@doi [Phys. Rev. D] {10.1103/PhysRevD.91.042003},
  91, 042003

\bibitem[\protect\citeauthoryear{{Weisstein}}{{Weisstein}}{2019}]{GammaWolfram}
{Weisstein} E.~W.,  2019, {Gamma Distribution}, From MathWorld--A Wolfram Web
  Resource. http://mathworld.wolfram.com/GammaDistribution.html

\bibitem[\protect\citeauthoryear{{Wong} et~al.,}{{Wong}
  et~al.}{2019}]{Holicow19}
{Wong} K.~C.,  et~al., 2019, arXiv e-prints, \href
  {https://ui.adsabs.harvard.edu/abs/2019arXiv190704869W} {p. arXiv:1907.04869}

\bibitem[\protect\citeauthoryear{Wu \& Huterer}{Wu \&
  Huterer}{2017}]{wu2017sample}
Wu H.-Y.,  Huterer D.,  2017, \mn@doi [MNRAS] {10.1093/mnras/stx1967}, 471,
  4946

\bibitem[\protect\citeauthoryear{Wyman, Rudd, Vanderveld  \& Hu}{Wyman
  et~al.}{2014}]{Wyman}
Wyman M.,  Rudd D.~H.,  Vanderveld R.~A.,   Hu W.,  2014, \mn@doi [Phys. Rev.
  Lett.] {10.1103/PhysRevLett.112.051302}, 112, 051302

\makeatother
\end{thebibliography}



\appendix

\section{{\tt PyCBC Inference} settings}
\label{sec:pycbc}
We infer the parameters of GW170817 within a Bayesian framework \citep{christensen_meyer}. The aim is to obtain the posterior probability density function for a set of parameters, {\boldmath$\theta$} $= [\theta_1, \theta_2, ..., \theta_n]$,  given the GW data, {\boldmath$x_{GW}$}(t) as follows:

\begin{equation}
\label{eq:bayes}
    p(\boldsymbol{\theta}|\boldsymbol{x_{GW}}(t), M )= \frac{p(\boldsymbol{\theta}|M) p(\boldsymbol{x_{GW}}(t)|\boldsymbol{\theta}, M)}{p(\boldsymbol{x_{GW}}(t)| M)}
\end{equation}
under the assumption of a GW model, $M$. $p(\boldsymbol{\theta} | M)$ is the prior probability on the waveform parameters which represents our knowledge about the parameters before considering the observations and  $p( \boldsymbol{x_{GW}}(t) | \boldsymbol{\theta}, M)$ is the likelihood of observing the data given the parameters. The term in the denominator is the evidence. We utilise the {\tt PyCBC Inference} software package \citep{biwer_pycbc, nitz_pycbc} with the parallel-tempered emcee sampler \citep{emcee} to perform the MCMC sampling.

The MCMC sampling is performed over the time of coalescence $t_c$, the GW polarisation angle $\psi$, the component spins $\chi_{1,2}$ , the luminosity distance $d_L$, the inclination angle of the binary $\iota$, the detector-frame chirp mass of the binary $\mathcal{M}^{det}$ and the mass ratio $q = m_1 / m_2 $ where $m_1$ is the primary mass and $m_2$ is the secondary mass with $m_1 > m_2$. The likelihood assumes a Gaussian model of detector noise and is marginalised over the phase $\phi$. Analytically marginalising over $\phi$ reduces the computation cost by a factor of $2-3$ \citep{biwer_pycbc}.

We assume a uniform prior on the time of coalescence in the GPS time interval $[1187008882.33, 1187008882.53]$ where the trigger time is $1187008882.43$. We use a uniform prior on the polarisation angle between $[0, 2\pi]$ and a uniform prior in  $\cos\iota$ for the inclination angle. A uniform prior is adopted for the component masses $m_1$ and $m_2$ in the interval $[1.0, 2.0]\ M_\odot$ and a uniform prior on the component spins between $[-0.05, 0.05]$ \citep{brown_bns_spin}. Finally we assume a uniform in volume prior on the distance in the interval $[5, 80]$ Mpc. We also fix the sky location of GW170817 to the location of the host galaxy NGC 4993 at R.A. $ = 197.450374 ^{\circ}$, dec $= -23.381495 ^{\circ}$ \citep{soares2017}. 

The waveform model used is the TaylorF2 aligned-spin waveform model \citep{BuonannoTF2}. The detector's noise power spectral density (PSD) used in the likelihood is constructed using the median PSD estimation method \citep{ALLEN_PSD_estimation} with $16$-second Hann-windowed segments overlapped by $8$ s and truncated to $8$ s in length in the time domain \citep{ALLEN_PSD_estimation}. A sample rate of  $2048$ Hz is used for the analysis. The GW data used in the likelihood are in the interval $1187008758$ to $1187008886$ and the likelihood is evaluated from a low-frequency cutoff of $20$ Hz.

We use the GW170817 strain data from the Advanced LIGO (Hanford and Livingston) and Virgo detectors made available through the LIGO Open Science Centre (LOSC) \citep{Vallisneri_losc}. We make use of the {\fontfamily{cmtt}\selectfont LOSC\_CLN\_4\_V1} data which include a post-processing noise subtraction.


\bsp	
\label{lastpage}
\end{document}